\documentclass[prb,amsmath,amssymb]{revtex4}
\usepackage{graphicx}

\begin{document}

\title{Magnetic field dependence of edge states in MoS$_2$ quantum dots}
\author{Carlos Segarra, Josep Planelles and Juan I. Climente}
\email{planelle@uji.es}
\affiliation{Departament de Qu\'{\i}mica F\'{\i}sica i Anal\'{\i}tica, Universitat Jaume I, 12080 Castell\'o, Spain}

\begin{abstract}
We study the electronic structure of monolayer MoS$_2$ quantum dots subject to a perpendicular magnetic field. 
The coupling between conduction and valence band gives rise to mid-gap topological states
which localize near the dot edge. These edge states are analogous to those of 1D quantum rings.
We show they present a large, Zeeman-like, linear splitting with the magnetic field,
anticross with the delocalized Fock-Darwin-like states of the dot,
give rise to Aharonov-Bohm-like oscillations of the conduction (valence) band low-lying states
in the K (K') valley, and modify the strong field Landau levels limit form of the energy spectrum.
\end{abstract}

\maketitle

Two-dimensional transition metal dichalcogenides (TMDs) have arised as an alternative to graphene for electronic 
and opto-electronic applications where a finite gap is required.\cite{KoperskiNP}
Recently, single photon emitter TMDs have been observed, whose quantum dot like behavior is typically 
associated with lattice defects.\cite{KoperskiNN,SrivastavaNN,HeNN,ChakrabortyNN,TonndorfOPT,KumarNL,BrannyAPL}
TMD quantum dots with controlled quantum confinement are now being pursued with different techniques,
including patterning of TMD monolayers\cite{WeiSR}, chemical synthesis\cite{LinACS,TranNN} and defect engineering\cite{TongaySR,ZhouNL}.
In this context, theoretical studies have arised investigating the electronic structure of TMD quantum dots. 
Particular interest has been placed in the response to external magnetic fields. 
Kormanyos and co-workers have analyzed the CB under perpendicular magnetic fields for hard wall circular Mo$S_2$ and W$S_2$ dots.\cite{KormanyosPRX}
The resulting spectrum is reminiscent of the Fock-Darwin spectrum in harmonically confined dots, but with sizable out-of-plane $g$ factors due to spin-orbit interaction. 
Brooks and Burkard showed that the magnetic field can be used to force spin degeneracies in spite of the spin-orbit splitting, which is of interest for development of spin qubits.\cite{BrooksPRB}
Dias and co-workers investigated the energy levels of CB and VB in $K$ and $K'$ valleys of MoS$_2$, as well as the associated magneto-absorption spectrum.\cite{DiasJPCM,QuSR}\\

The above studies, however, have not considered the possible presence of edge states, which show up in the gap  of finite MoS$_2$ systems under different conditions.\cite{BollingerPRL,PanJMC,ErdoganEPJB,PavlovicPRB,DavelouSSC,PeterfalviPRB,SegarraPRB} The origin of such states lies in the marginal topological properties of the single-valley MoS$_2$ Hamiltonian. In k$\cdot$p formalism\cite{Kormanyos2D}, these properties manifest when one expands the Hamiltonian up to second order in $k$ and explicitly considers the CB-VB coupling.\cite{SegarraPRB}
In this chapter, we analyze the response of mid-gap monolayer MoS$_2$ quantum dot states to 
perpendicular magnetic fields. 
To this end, we use a two-band k$\cdot$p Hamiltonian:
\begin{equation}
H = \left( 
\begin{array}{cc}
E_v + \alpha \, p^2 - V(\mathbf{r}) & \tau \gamma p_- \\
\tau \gamma p_+ & E_c + \beta \, p^2 + V(\mathbf{r}) 
\end{array}
\right).
\label{eq:H_MoS2}
\end{equation}
\noindent where $p_\pm = p_x \pm i \tau p_y$ and $\mathbf{p}=\mathbf{k} + \mathbf{A}$, with $\mathbf{k}$ the momentum relative to the $K/K'$ points and $\mathbf{A}=B/2\,(-y,x,0)$ the vector potential. 
$B$ is the magnetic field, $E_c = \Delta/2$ and $E_v= -\Delta/2$ the CB and VB edge energies, respectively, $\Delta$ is the band gap.
The constants $\alpha$, $\beta$ and $\gamma$ are material parameters, while $\tau$ identifies the valley $K$ ($\tau = 1$) or $K'$ ($\tau=-1$). 
$V(\mathbf{r})$ represents a possibly externally applied potential as e.g. electrostatic gating. If no external potential is present, then $V(\mathbf{r})=0$.
We impose hard-wall potential at the QD border, the associated boundary conditions result in no intervalley coupling.
Notice also that for clarity we ignore spin and spin-orbit terms, which in MoS$_2$ give rise to small energy splittings of levels at zero and finite magnetic fields.\cite{DiasJPCM}
We also disregard trigonal warping and other minor corrections to the Hamiltonian.\cite{Kormanyos2D}
Hamiltonian ($\ref{eq:H_MoS2}$) is solved numerically.

\begin{figure}[h]
\includegraphics[scale=.6]{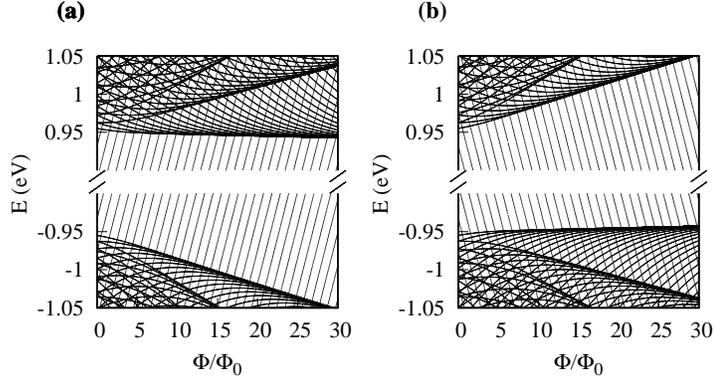}
\caption{Energy levels of circular dots as a function of a perpendicular magnetic flux, for $\alpha=-\beta=1$ eV$\cdot$\AA$^2$.
(a): $K$ valley. (b) $K'$ valley.}\label{fig9}
\end{figure}

To illustrate the effect of the magnetic field on the electronic structure, we first consider the highly symmetric case of circular quantum dots with 
equivalent masses in CB and VB, $\alpha=1$ eV$\cdot$\AA$^2$ and $\beta=-1$ eV$\cdot$\AA$^2$, along with other MoS$_2$ parameters ($\gamma=3.82$ eV$\cdot$\AA, $\Delta=1.9$ eV) 
and radius $R=9$ nm. 
Fig.~\ref{fig9} (a) and (b) show the energy levels in the $K$ and $K'$ valley, respectively, as a function of the magnetic flux $\Phi = B S/\Phi_0$,
with $S$ the dot surface and $\Phi_0=2\pi$ the unit quantum flux (in atomic units).
As can be seen, CB ($E > 0.95$ eV) and VB ($E < -0.95$ eV) display a Fock-Darwin like spectra, 
where spatially confined states converge into Landau levels (LLs) with increasing flux. 
Notice the LLs of 2D TMDs include energy-locked levels which are independent of $\Phi$,
as can be seen in the lowest level of the CB of $K$, in Fig.~\ref{fig9}(a). 
Besides, CB of $K$ ($K'$) valley and VB of $K'$ ($K$) valley are mirror images.
Up to this point, all features are consistent with the picture described by Dias et al.\cite{DiasJPCM,QuSR}

However, superimposed to the Fock-Darwin like spectrum, there are a series of iso-spaced states which show a identical linear 
dispersion with the field, covering the entire spectrum: CB, VB and gap region alike.
These are the edge states of the dot, arising from the marginal topological character of Hamiltonian ($\ref{eq:H_MoS2}$).\cite{SegarraPRB} 

The slope of edge states against $\Phi$ is positive for $K$ and negative for $K'$ valleys, evidencing a large Zeeman level splitting.
 The sign and magnitude can be understood by simplifying Hamiltonian ($\ref{eq:H_MoS2}$) for a circular structure and fixing the radius to $R$, as expected for pure edge states.
The resulting Hamiltonian, neglecting magnetic field for the moment, is:
\begin{equation}
\label{eq:ham3}
H_R=
 \left(
\begin{array}{cc}
 \varepsilon_v +  \frac{\alpha}{R^2} \hat L_z^2 & -i \, \frac{\tau \gamma}{R} \, e^{-i \theta} \, \hat L_z \\
 i \,\frac{\tau \gamma}{R} \, e^{i \theta} \hat L_z & \varepsilon_c + \frac{\beta}{R^2} \, \hat L_z^2 
\end{array}
\right).
\end{equation}
\noindent with $\hat L_z$ the azimuthal angular momentum operator. 
The  eigenvectors are spinors $\Psi=\left( a \, e^{i M \theta}, b \, e^{i (M+1) \theta} \right)$, with $a^2=\frac{|\beta|}{|\alpha|+|\beta|}$, $b^2=\frac{|\alpha|}{|\alpha|+|\beta|}$,\cite{SegarraPRB} and $M$ is the $\hat L_z$ quantum number. The mean value of the energy, $E=\langle \Psi | H_R | \Psi \rangle$, is
\begin{equation}
\label{eq:ene}
E = 
     (|a|^2 \varepsilon_v+ |b|^2 \varepsilon_c)+\frac{i \, \tau \gamma}{R} (M \,(a b^*-a^* b)-a^* b)+ \frac{|a|^2 \alpha}{R^2} \, M^2 + \frac{|b|^2 \beta}{R^2} \, (M+1)^2.
\end{equation}

\noindent Since $E$ must be real, so must be $i\, a^* b$ and $i\,(a b^*-a^* b)$.
Therefore, one of the two complex constants must be a real number and the other one an imaginary number. 
Let us assume $a$ is real and $b$ a pure imaginary number. Then:
\begin{equation}
\label{eq:ene1}
E 
  =  (|a|^2 \varepsilon_v+ |b|^2 \varepsilon_c)+\frac{M^2}{R^2}(\alpha |a|^2+\beta |b|^2)+ (2 M+1) \frac{\beta |b|^2}{R^2}+\frac{\tau \gamma}{R} \left( 2 a |b| M + a |b|\right).
\end{equation}
Furthermore, since $\alpha>0$, $\beta<0$, $a^2=\frac{|\beta|}{|\alpha|+|\beta|}$ and $b^2=\frac{|\alpha|}{|\alpha|+|\beta|}$, we have that $(\alpha |a|^2+\beta |b|^2)=0$.
Thus,
\begin{eqnarray}
\label{eq:ene2}
E &=& (\frac{|\beta|}{|\alpha|+|\beta|} \varepsilon_v+ \frac{|\alpha|}{|\alpha|+|\beta|} \varepsilon_c)+\frac{\beta \alpha}{|\alpha|+|\beta|}\frac{1}{R^2}+
\frac{2 \beta \alpha}{|\alpha|+|\beta|}\frac{M}{R^2} \nonumber \\
&+& \frac{2 \tau \gamma \sqrt{\alpha |\beta|}}{(|\alpha|+|\beta|) R} \, M -
\frac{\tau \gamma \sqrt{\alpha |\beta|}}{(|\alpha|+|\beta|) R}.
\end{eqnarray}
\noindent The presence of a magnetic flux $\Phi$ can be incorporated by the formal replacement $M\to M+\Phi$,
 so that the flux-dependent energy $E_{\Phi}$ results in a linear dependence on the magnetic flux:
\begin{equation}
\label{eq:ene3}
E_{\Phi}=E+\left(\frac{2 \beta \alpha}{|\alpha|+|\beta|}\frac{1}{R^2}+\frac{2 \tau \gamma \sqrt{\alpha |\beta|}}{(|\alpha|+|\beta|)}\frac{1}{R} \right) \, \Phi
\end{equation}
\noindent For the particular case of $|\alpha|=|\beta|$, i.e., $a^2=b^2=1/2$, the slope of the flux becomes  $(\frac{\sqrt{|\beta|\alpha}}{R^2} + \frac{\tau \gamma}{R})$,
which is in quantitative agreement with the slope of the edge states numerically calculated and shown in Fig.~\ref{fig9}. We note the second term in the slope, arising from the off-diagonal band coupling
in Hamiltonian ($\ref{eq:H_MoS2}$), is the dominant term, which explains the opposite slope in $K$ ($\tau=1$) and $K'$ ($\tau=-1$) valleys.\\

\begin{figure}[h]
\includegraphics[scale=.6]{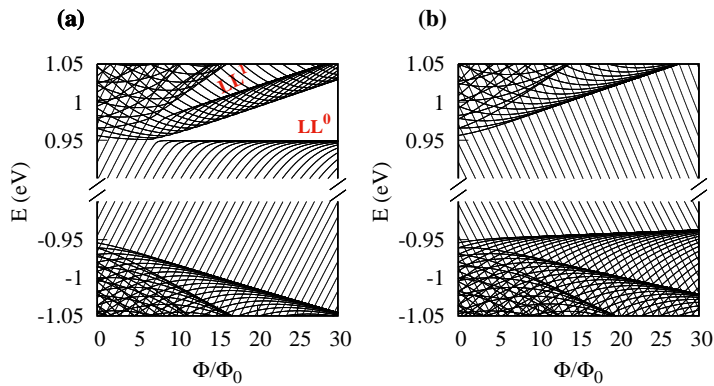}
\caption{Energy levels of a MoS$_2$ circular dot as a function of a perpendicular magnetic flux. $\alpha=1.72$ eV$\cdot$\AA$^2$
and $\beta=-0.13$ eV$\cdot$\AA$^2$.  (a): $K$ valley. (b) $K'$ valley. The lowest (LL$^0$) and first excited (LL$^1$) Landau levels of the CB are labeled.}\label{fig10}
\end{figure}

Considering the pervasive presence of edge states in the magneto-spectrum of Fig.~\ref{fig9}, one suspects they could have important implications
for actual magneto-absorption and spin properties of TMD dots. Since early theoretical studies overlooked such states,\cite{DiasJPCM,QuSR,BrooksPRB}
 next we explore their robutstness when using actual MoS$_2$ mass parameters, $\alpha=1.72$ eV$\cdot$\AA$^2$ and $\beta=-0.13$ eV$\cdot$\AA$^2$.\cite{KormanyosPRB}
The results are shown in Fig.~\ref{fig10}. As can be seen, the spectra are similar to those of Fig.~\ref{fig9}, except for the CB of the $K$ point 
--see top of Fig.~\ref{fig10}(a)--, where drastic changes appear. Here, a gap opens up between the lowest ($\Phi$-independent) LL and higher states,
and Aharonov-Bohm like oscillations take place in the many-fold of states under each excited LL. 

The interpretation of these effects is as follows. For $|\alpha|=|\beta|$ the Fermi level was in the center of the gap, $E \approx 0$ (notice the summands in the first parenthesis of Eq.(\ref{eq:ene2}) cancel out), and so were the edge states with small $M$ angular momentum. Instead, for $|\alpha| \gg |\beta|$, the Fermi level shifts towards the vicinity of the CB. In the $K$ valley, where such states have positive slope, this enables anticrossings between edge states and corresponding CB states with the same $M$. No anticrossings are observed in the $K'$ valley because the low-$M$ edge states, being close to the CB at zero field, require stronger $\Phi$ than we show in Fig.~\ref{fig10} to reach their VB counterparts.

\begin{figure}[h]
\includegraphics[scale=.6]{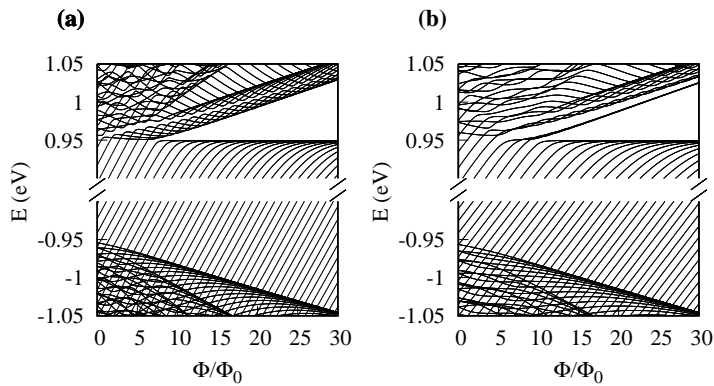}
\caption{Same as Fig.~\ref{fig10}(a), but for hexagonal (a) and triangular (b) dots.}\label{fig11} 
\end{figure}

Similar results are obtained if circular confinement is replaced by other shapes. Fig.~\ref{fig11} (a) and (b) show the magneto-spectrum of hexagonal and triangular MoS$_2$ quantum dots, respectively. Edge states again anticross with CB states, opening gaps and forming Aharonov-Bohm like oscillations. The main difference as compared to circular dots is that the oscillating many-folds are now formed by sets of six (Fig.~\ref{fig11}(a)) and three (Fig.~\ref{fig11}(b)) energy levels. This is due to the reduced symmetry of hexagons ($C_6$) and triangles ($C_3$) as compared to the circle.

\begin{figure}[h]
\includegraphics[scale=.4]{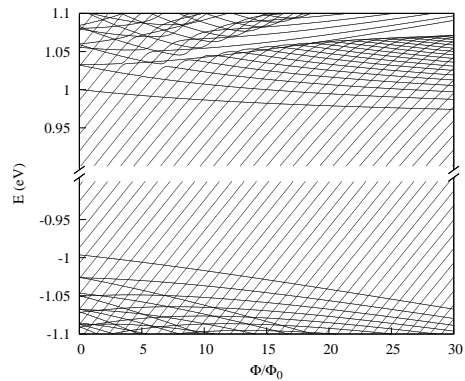}
\caption{Same as Fig.~\ref{fig10}(a), but including a harmonic confinement potential, $V(r)=1/2\,m_j\,\omega_j^2\,r^2$ ($j=e,h$). $m_e=1/\beta$, $m_h=1/\alpha$, $\omega_e=30$ meV and $\omega_h=\omega_e\,\sqrt{m_e/m_h}$.}\label{fig12}
\end{figure}

The quantum ring-like behavior of MoS$_2$ quantum dots arising from edge states can be tailored by means of external fields. As an example, in Fig.~\ref{fig12} we represent the $K$ valley of a circular dot like that in Fig.~\ref{fig10}(a), but adding a harmonic confinement potential, which could be associated e.g. to electrostatic gating, $V(r)=1/2\,k\,r^2$, with $k=m_j\,\omega_j^2$ ($j=CB,VB$ for electrons and holes). Edge states turn out to be robust against such potential, which is quite strong near the boundaries, see Fig.~\ref{fig12}, but they are energetically unstabilized. In particular, low $M$ edge states are blueshifted away from the middle of the gap, towards the proximity of the CB. This change shifts anticrossings with CB states to weaker $\Phi$ values as compared to the system with $V(r)=0$, Fig.~\ref{fig10}(a). Consequently, anticrossings take place in excited CB states. 

In conclusion, edge states in monolayer TMDs quantum dots exhibit a linear, Zeeman-like, response against perpendicular magnetic fields. When anticrossing with delocalized states of the dot, they can give rise to Aharonov-Bohm like oscillations. For MoS$_2$ quantum dots, these features are expected to show up in the CB of the $K$ valley (and, for stronger fields, in the VB of the $K'$ valley). The addition of external potentials, modifying the edge states energy with respect to that of delocalized states, can be used to tune the magnetic fields at which these quantum ring like features takes place.

\begin{acknowledgments}
Support from MICINN project CTQ2014-60178-P and UJI project P1-1B2014-24 is acknowledged.
\end{acknowledgments}

\end{document}